\documentclass[hyper,11pt,letterpaper]{JHEP3}
\usepackage[dvips]{epsfig}
\usepackage{amsmath,amssymb,epsf}
\usepackage{cite}
\usepackage{graphicx}
\usepackage{dcolumn}
\usepackage{bm}

\newcommand{\be}{\begin{equation}}
\newcommand{\ee}{\end{equation}}
\newcommand{\bea}{\begin{eqnarray}}
\newcommand{\eea}{\end{eqnarray}}
\newcommand{\ba}{\begin{eqnarray}}
\newcommand{\ea}{\end{eqnarray}}

\newcommand{\beq}{\begin{equation}}
\newcommand{\eeq}{\end{equation}}
\newcommand{\beqa}{\begin{eqnarray}}
\newcommand{\eeqa}{\end{eqnarray}}
\newcommand{\beqar}{\begin{eqnarray*}}
\newcommand{\eeqar}{\end{eqnarray*}}

\newcommand{\noi}{\noindent}

\newcommand{\N}{{\mathcal{N}}}
\newcommand{\R}{{\mathbb{R}}}

\def\a{\alpha}

\def\m{\mu}

\def\p{\pi}                

\def\G{\Gamma}

\def\Nc{N_c}
\def\Nf{N_f}

\def\Nfour{\mathcal N\,{=}\,4}
\def\Ntwo{\mathcal N\,{=}\,2}

\def\Om{{\cal{O}}_m}
\def\la{\langle}
\def\ra{\rangle}
\def\rarrow{\rightarrow}

\title{\LARGE Holographic Thermodynamics at Finite Baryon Density: Some Exact Results}

\author{Andreas Karch\footnotemark[1]\,
and Andy O'Bannon,\footnotemark[2]\,
\\
Department of Physics, University of Washington, Seattle, WA 98195-1560}

\footnotetext[1]{E-mail: \email{karch@phys.washington.edu}}
\footnotetext[2]{E-mail: \email{ahob@u.washington.edu}}

\abstract{We use the AdS/CFT correspondence to study the thermodynamics of  massive $\N=2$ supersymmetric hypermultiplets
coupled to $\N=4$ supersymmetric $SU(N_c)$ Yang-Mills theory in the limits of large $N_c$ and large 't Hooft coupling. In particular, we study the theory at finite baryon number density. At zero temperature, we present an exact expression for the hypermultiplets' leading-order contribution to the free energy, and in the supergravity description we clarify which D-brane configuration is appropriate for any given value of the chemical potential. We find a second-order phase transition when the chemical potential equals the mass. At finite temperature, we present an exact expression for the hypermultiplets' leading-order contribution to the free energy at zero mass.}


\begin{document}

\section{Introduction and Review}  \label{intro}

Understanding the phase diagram of Quantum Chromodynamics (QCD) is
essential for understanding a variety of physical systems, such as
the early universe and neutron stars. Much of the interesting thermodynamics of QCD, such as the deconfinement transition, occurs when traditional perturbative methods are unreliable. The lattice formulation of the theory circumvents this problem by direct calculation of the path integral. Implementing a baryon number
chemical potential on the lattice is difficult, however. We are thus led to study the thermodynamics of model theories similar to QCD, but that offer more tractable
strong-coupling analysis. In particular, the AdS/CFT correspondence,
and more generally gauge-gravity duality, permits quantitative study
of the thermodynamics of strongly-coupled non-Abelian gauge theories
that share some features with QCD.

In this paper we will use the AdS/CFT correspondence to study the
$\N=4$ supersymmetric $SU(N_c)$ Yang-Mills (SYM) theory, in the 't
Hooft limit of large-$\Nc$, when the 't Hooft coupling is large,
$\lambda \equiv g_{YM}^2 \Nc \gg 1$. The $\Nfour$ SYM theory is very
different from QCD at zero temperature, being a conformal field
theory (CFT) and hence lacking confinement, a mass gap, and chiral
symmetry breaking. At finite temperature, however, the $\Nfour$ SYM
theory resembles the high-temperature phase of QCD, in particular by
exhibiting screening of color charges. The $\Nfour$ SYM theory at
finite temperature has no phase transitions, however. Being a CFT,
the theory has no intrinsic scale to set a transition temperature,
so no transitions can occur.

The AdS/CFT correspondence is the conjecture that the $\Nfour$ SYM theory,
in the limits described above, is equivalent to type IIB
supergravity formulated on the background spacetime $AdS_5 \times
S^5$, where $AdS_5$ is five-dimensional anti-de Sitter space and
$S^5$ is a five-sphere \cite{jthroat,EW,GKP}. This spacetime arises
as the near-horizon geometry of $\Nc \rarrow \infty$ coincident
D3-branes in type IIB string theory. Thermal equilibrium of the
$\Nfour$ SYM theory corresponds to non-extremal D3-branes, whose
near-horizon geometry is five-dimensional AdS-Schwarzschild times
$S^5$ \cite{Witten:1998zw,Gubser:1996de}. The temperature $T$ of the
$\Nfour$ SYM theory is identified with the Hawking temperature of
the AdS-Schwarzschild black hole.

The $\Nfour$ SYM theory, unlike QCD, only describes fields in the
adjoint representation of the gauge group. We will introduce fields
in the fundamental representation in the form of $\Ntwo$
supersymmetric hypermultiplet ``flavor'' fields. We will introduce a
finite number $\Nf$ of them, so that $\Nf \ll \Nc$ in the 't Hooft
limit, and work to leading order in $\Nf / \Nc$. To this order, the
beta function of the theory is zero and hence the theory remains
(approximately) conformal. We may introduce an $\Ntwo$
supersymmetry-preserving mass $m$ for the hypermultiplets, however,
which of course explicitly breaks the conformal invariance.

The $\Nf$ hypermultiplets possess a $U(N_f)$ global symmetry
analogous to the vector flavor symmetry of QCD. As in QCD, we
identify the $U(1)_B$ subgroup of $U(N_f)$ as baryon number. We will
introduce a finite $U(1)_B$ density $\la J^t \ra$, where $J^{\mu}$
is the $U(1)_B$ current, or equivalently we work with a finite
chemical potential $\mu$.

The flavor fields appear in the supergravity description as $\Nf$
D7-branes \cite{KarchKatz}. In the $\Nf \ll \Nc$ limit, the
contribution that these D7-branes make to the stress-energy tensor
is dwarfed by the contribution of the $\Nc$ D3-branes. To leading order in $\Nf / \Nc$, then, we may neglect the back-reaction of the
D7-branes on the geometry: they are probes. The D7-brane action is
then the Dirac-Born-Infeld (DBI) action. The hypermultiplet mass $m$
appears in the supergravity description as the asymptotic separation
of the D3-branes and D7-branes in a mutually orthogonal direction.
The $U(N_f)$ flavor symmetry appears in the supergravity description
as the $U(N_f)$ worldvolume gauge invariance of the $\Nf$ D7-branes.
The $U(1)_B$ current $J^{\mu}$ is dual to the $U(1)$ worldvolume
gauge field $A_{\mu}$. To obtain a finite baryon number density $\la
J^t \ra$ in the SYM theory, we must introduce a nonzero gauge field
time component, $A_t(r)$, where $r$ is the radial coordinate of AdS.

The SYM theory has three dimensionful parameters: the temperature, $T$,
the hypermultiplet mass $m$, and the baryon number density, $\la J^t \ra$. We
will use the D7-brane description of the flavor fields to derive
exact expressions for their free energy in two
limits: $T=0$ and $m=0$. To place our results in context, we will
now review the known thermodynamics of the flavor fields in the SYM
theory, as deduced from supergravity calculations.

Consider first zero temperature, zero mass, and zero density in the
SYM theory. The D3-branes and D7-branes then have zero asymptotic
separation. The D7-branes are extended along $AdS_5 \times S^3$
inside of $AdS_5 \times S^5$, where the worldvolume $S^3$ is the maximum-volume equatorial $S^3 \subset S^5$ \cite{KarchKatz}. We first turn on $m$, which means
finite asymptotic separation of the D3-branes and D7-branes. The
D7-brane then has a non-trivial embedding in the near-horizon
geometry. The D7-brane wraps the equatorial $S^3 \subset S^5$ at the boundary of $AdS_5$. Moving into the bulk of
$AdS_5$, the position of the $S^3$ can change, that is, the D7-brane
``slips'' on the $S^5$. The volume of the $S^3$ shrinks as we move
into $AdS_5$ and indeed may ultimately vanish at some radial
position $r'$, as allowed by topology (the $S^3$ is a trivial
cycle). The D7-brane then does not extend past $r'$: it appears to
``end'' at $r'$ \cite{KarchKatz}. The worldvolume scalar that describes the embedding
is dual holographically to an operator $\Om$ given by taking
$\frac{\partial}{\partial m}$ of the SYM Lagrangian. In particular,
$\Om$ includes the mass operator of the fermions in the
hypermultiplet as well as $m$ times the mass operator of the
scalars, as well as couplings to adjoint scalars. The exact operator
is written in ref. \cite{Kobayashi:2006sb}. For our purposes,
thinking of $\Om$ as the mass operator will be sufficient.

Now turn on the temperature $T$. The spacetime is now
AdS-Schwarzschild, and a horizon is present at some radial position
$r_H$. Two classes of embedding are now possible. The first are the
finite-temperature analogues of the embeddings described above:
D7-branes that end outside the horizon, \textit{i.e.} $r' > r_H$.
These are called ``Minkowski'' embeddings. The second class are
D7-branes in which the $S^3 \subset S^5$ shrinks as we move away
from the boundary, but never collapses to zero volume. The D7-brane
then extends to, and intersects, the AdS-Schwarzschild horizon.
These are called ``black hole'' embeddings. A topology-changing
phase transition exists between these two types of embeddings, which
appears in the SYM theory as a first-order phase transition
associated with the flavor fields as we change $m/T$
\cite{Babington:2003vm,Kirsch:2004km,Ghoroku:2005tf,Apreda:2005yz,Mateos:2006nu,Albash:2006ew,andy,Mateos:2007vn}.

Now turn on the baryon number density $\la J^t \ra$. The $U(1)$ gauge field on
the D7-brane worldvolume now has nonzero $A_t(r)$. The asymptotic
value $A_t(\infty)$ gives the value of $\mu$. In the canonical
ensemble, the topology-changing transition observed at zero density
persists for small density and then ends in a critical point
\cite{Kobayashi:2006sb}. A region of instability exists, with the
line of first-order transitions as a boundary
\cite{Kobayashi:2006sb}.

A claim was made in ref. \cite{Kobayashi:2006sb} that when $F_{tr}$ is nonzero only black
hole embeddings are physically allowed. Minkowski embeddings were dismissed
as unphysical because, if the D7-brane ends outside the horizon, the
radial electric field lines $F_{tr}$ have no place to end. If a source, a
density $\la J^t \ra$ of fundamental strings stretching from the D7-brane to
the horizon, is introduced to accommodate the field lines, the
tension of the strings will overcome the tension of the D7-brane,
drawing the D7-brane into the horizon and producing a black hole
embedding. Minkowski embeddings were thus excluded from the analysis
of ref. \cite{Kobayashi:2006sb}.

In refs. \cite{Nakamura:2006xk,Nakamura:2007nx}, however, an argument was made that Minkowski embeddings were physical and should be included. One of the main objections to the claim of ref. \cite{Kobayashi:2006sb} was that the black hole
configurations cannot realize chemical potentials below a critical value, which at zero temperature we will show is $\mu=m$. What we find is that this ``incompleteness problem'' is completely resolved by including Minkowski branes with a \textit{constant} $A_t(r)$, that is, with non-vanishing chemical potential but zero density\footnote{While this paper was in preparation ref. \cite{Ghoroku:2007re} appeared, where the same conclusion was reached.}. The authors of refs. \cite{Nakamura:2006xk,Nakamura:2007nx} argue that $A_t$ is not a gauge-invariant parameter, but in the thermodynamic setting where spacetime has a compact Euclidean time direction, the Wilson line of $A_t$ is gauge invariant and hence physical. The constant part in $A_t(r)$ is the chemical potential according to the standard rules of AdS/CFT, since it represents the leading near-boundary behavior. For any configuration that satisfies the boundary condition that $A_t(r)$ vanish in the infrared (deep in AdS), this definition becomes equivalent to the description $\mu = \int F_{rt}$ advertised in refs. \cite{Nakamura:2006xk,Nakamura:2007nx}. The latter definition misses the important case that $A_t(r)$ is simply constant, however, which is always a solution in the absence of sources.

We will find analytic solutions for the zero-temperature limit of black hole embeddings, and will argue that these are thermodynamically favored when $\mu > m$ and $\la J^t \ra >0$. This is in complete agreement with the claims of ref. \cite{Kobayashi:2006sb}, where only $\la J^t \ra > 0$ was considered. When $\mu \leq m$, Minkowski embeddings with constant $A_t(r)$, and hence $\la J^t \ra=0$, are thermodynamically favored. The transition between the two is second order. Indeed, when $\mu > m$ we see, as expected, a free energy whose form is consistent with spontaneous symmetry breaking for a scalar field with a wine-bottle potential. When $\mu = m$ the minimum of the scalar's potential returns to the origin. Note in particular that we will not see a Fermi surface: the system prefers to condense scalars.

We can also generalize our zero-temperature analysis to include nonzero currents $\la J^i \ra$, for spatial index $i$. At zero temperature, these currents will not dissipate and can be introduced as a constant background. We can generalize further, from a D7-brane probe to a D5-brane probe \cite{Karch:2000gx,DeWolfe:2001pq}. Indeed, our analysis should generalize to many probe D-brane systems.

At finite temperature, we are also able to provide an analytic expression for the hypermultiplets' contribution to the free energy at leading order in $N_f / N_c$ in the limit $m = 0$. The result exhibits no interesting phase structure, in accord with the phase diagram of ref. \cite{Kobayashi:2006sb}.

Another holographic model of QCD at finite baryon density, the D4/D8 system, and specifically the Sakai-Sugimoto model \cite{Sakai:2004cn}, has received a great deal of attention recently
\cite{Aharony:2006da,Kim:2006gp,Horigome:2006xu,Parnachev:2006ev,Yamada:2007ys,Bergman:2007wp,Davis:2007ka,Rozali:2007rx}.
In this case, the supergravity background is provided by D4-branes and the flavor branes are D8-branes and anti-D8-branes. As in ref. \cite{Herzog:2007kh}, in the course of our analysis we will find a D7-brane configuration similar to the D8-brane in the Sakai-Sugimoto model, but, also as in ref. \cite{Herzog:2007kh}, we will find that these configurations are always thermodynamically disfavored. Indeed, these solutions should not be compared to any of our other solutions, as they are non-supersymmetric even at zero density and describe a SYM Lagrangian distinctly different from that of the SYM theory described above.

Most of the work on the Sakai-Sugimoto model at finite density
is based on analogues of Minkowski embeddings, with strings attached (or
alternatively a wrapped D4-brane providing the required source). Our
results, as well as the earlier work of ref.
\cite{Kobayashi:2006sb}, show that in the analytically tractable
D3/D7 system, such configurations are not solutions of the full
source-plus-brane system, and do not even provide a good qualitative
picture of the phase structure. This at least suggests that any
analysis based on approximating a full solution to the non-abelian
DBI action relevant in the Sakai-Sugimoto model with a source-plus-brane
configuration has to be approached with some caution. Presumably a
solution like the smooth black hole embedding for the D7-brane
discussed here exists in that case as well, but it most likely will
involve non-abelian configurations of scalars and gauge fields as well
as the tachyon.

This paper is organized as follows. In section \ref{zeroT} we
present our zero-temperature analysis. In section \ref{zeromass} we present our zero mass analysis. We conclude in section \ref{conclusion}. We collect various necessary integrals in the Appendix.

\section{The Limit of Zero Temperature}  \label{zeroT}

\subsection{Analytic Brane Embeddings}  \label{anaemb}

In type IIB string theory, we consider a system of $N_c$ D3-branes and $N_f$ D7-branes aligned in flat ten-dimensional space as

\begin{equation}
\begin{array}{ccccccccccc}
   & x_0 & x_1 & x_2 & x_3 & x_4 & x_5 & x_6 & x_7 & x_8 & x_9\\
\mbox{D3} & \times & \times & \times & \times & & &  &  & & \\
\mbox{D7} & \times & \times & \times & \times & \times  & \times
& \times & \times &  &   \\
\end{array}
\end{equation}

\noi If we separate the D3-branes and D7-branes in a mutually
orthogonal direction, in the $x_8$-$x_9$ plane, an open string
stretched between them will have a mass equal to the separation
times the string tension. This mass appears as the hypermultiplet
mass $m$ in the D3-brane worldvolume theory. We take the usual
AdS/CFT limit, $N_c \rarrow \infty$, $g_s \rarrow 0$ with $g_s N_c$
fixed and $g_s N_c \gg 1$ \cite{jthroat}. We obtain the near-horizon
geometry of D3-branes, $AdS_5 \times S^5$, with the $N_f \ll N_c$
D7-branes extended along $AdS_5 \times S^3$.

The DBI action describing the embedding of any probe D-brane is
non-linear. Even when the symmetries are strong enough to restrict
all the embedding functions to depend on a single variable only, the
resulting ordinary differential equation of motion can often only be
solved numerically. Sometimes symmetries provide integrals of motion
that permit an exact solution of the system. For example, in the
Sakai-Sugimoto model \cite{Sakai:2004cn} translation invariance
along the direction $x$ transverse to the D8-branes ensures that the D8-brane
action only depends on derivatives of $x$ and not on $x$ itself. We
will show that the same is true for the D3/D7 system at zero
temperature even in the presence of certain worldvolume gauge
fields. From the SYM perspective, this is the statement that we may
determine the free energy exactly for non-vanishing currents $J^{\mu}$, and in particular for finite
density $\la J^t \ra$.

In the original analysis of ref. \cite{KarchKatz}, the embedding of the
D7-brane in AdS$_5 \times S^5$ was
characterized by a slipping mode $\theta(r_6)$. That is, the AdS$_5$
$\times$ $S^5$ metric was written as

\beq
\label{kkmetric} ds^2 = \frac{r_6^2}{R^2} \eta_{\mu \nu}dx^{\mu}
dx^{\nu} + \frac{R^2}{r_6^2} dr_6^2 + (d \theta^2 + \sin^2\theta \,
ds^2_{S^1} + \cos^2 \theta \, ds^2_{S^3})
\eeq

\noi where $\mu,\nu=0,1,2,3$ denote the four worldvolume directions
of the D3-brane, $\eta_{\mu \nu}$ is the (3+1)-dimensional Minkowski
metric, $R$ is the curvature radius\footnote{Starting now, we work
in units with $R \equiv 1$. In particular, we convert between
supergravity and SYM quantities with $\a' = \lambda^{-1/2}$.},
$ds^2_{S^1}$ and $ds^2_{S^3}$ are the metrics of unit 1- and
3-spheres, respectively, and $r_6$ is the radial
direction in the full space transverse to the D3-branes. In terms of
the ten-dimensional space in which the $N_c$-D3 branes are embedded,
$r_6^2 = \sum_a x_a^2$, where $a$ runs from 4 to 9. The action
of a D7-brane wrapping AdS$_5 \times S^3$ is proportional to the
volume of the $S^3$, which goes as $\cos^3\theta$, so we will find a
non-trivial potential for the embedding scalar $\theta(r_6)$ in this
case, and no simple integral of motion is apparent. However, if
we write the background metric as

\beq
\label{warpfactormetric}
ds^2 = Z(r_6) \eta_{\mu \nu}dx^{\mu} dx^{\nu} + Z^{-1}(r_6) (dr^2 + r^2 ds^2_{S^3} + dy^2 + dz^2),
\eeq

\noi a significant simplification occurs. In this case, $y(r)$ and
$z(r)$, which stand for $x_8$ and $x_9$, are the scalars describing
the embedding, and without loss of generality we can use the $U(1)$
symmetry that rotates them into one another to set $z=0$. $r$ is a
radial coordinate in four of the six transverse directions, so we
have $r_6^2 = r^2 + y^2+z^2$. All the $y$-dependence of the action appears via $r_6$ in the warp factors $Z(r_6)$. For the
embeddings we are interested in, with a D7-brane wrapping the four
$x_{\mu}$ directions and four of the six transverse directions, all
the warp factors will drop out from the action, and we are effectively
solving for D7-brane embeddings in flat space. The same simplification occurs in all supersymmetric probe brane embeddings where background and probe branes have four
relative ND directions. The action now only depends on derivatives
of $y$, and the resulting constant of motion is sufficient to
integrate analytically the equations of motion. This cancellation of warp factors is spoiled at finite temperature or in the presence
of electric or magnetic background fields in the SYM theory. This
coordinate system was originally used in ref. \cite{myers} to find
the exact meson spectrum of the SYM theory from fluctuations of the
D7-brane. More recently, it was used in ref. \cite{Herzog:2007kh} to
find new non-supersymmetric D7/anti-D7-brane embeddings. To
translate back to the original $\theta(r)$ embedding scalar, we can
always combine the definitions $y=r_6 \sin \theta$ and (at $z=0$)
$r_6^2 = r^2 + y^2$ to obtain $\tan \theta = \frac{y(r)}{r}$.

The action for the D7-branes is

\beq
S_{D7} = - N_f T_{D7} \int d^8 \xi \sqrt{-det (g_{ab} + (2\p\a') F_{ab})}
\label{dbi}
\eeq

\noi where $T_{D7}$ is the D7-brane tension, $\xi_a$ are worldvolume
coordinates, $g_{ab}$ is the induced worldvolume metric and $F_{ab}$
is the worldvolume $U(1)$ gauge field. Introducing the field
strength\footnote{Throughout this paper, we work in a gauge with
$A_r = 0$.} $F_{rt} = A_t'(r)$ and the embedding scalar $y(r)$, the
zero-temperature D7-brane Lagrangian is

\beq
L = -\N r^3 \sqrt{1+ y'^2 - (2\p\a')^2 A_t'^2}.
\eeq

\noi where $\N = N_f T_{D7} V_3$ with $V_3 = 2\p^2$ the volume of a
unit $S^3$. In terms of SYM theory quantities,

\beq
\N = \frac{\lambda}{(2\p)^4} N_f N_c
\label{dbiprefactor}
\eeq

\noi We will divide both sides of eq. (\ref{dbi}) by the volume of $\R^{3,1}$ and henceforth work with action (and in a moment, free energy) densities, so that $S_{D7} = \int dr L$.

Thanks to the warp factor cancellation, only derivatives of $y(r)$
and $A_t(r)$ appear in the Lagrangian, and we find two conserved
charges

\beq
\frac{\delta L}{\delta y'} = - \N r^3 \frac{y'}{\sqrt{1+ y'^2 -
(2\p\a')^2 A_t'^2}} \equiv -c, \qquad \frac{\delta L}{\delta A_t'} =
\N r^3 \frac{(2\p\a')^2 A_t'}{\sqrt{1+ y'^2 - (2\p\a')^2 A_t'^2}}
\equiv d.
\eeq

\noi The ratio implies

\beq
A_t'^2 = \frac{d^2}{(2\p\a')^4 c^2} y'^2.
\label{ratio}
\eeq

\noi We can algebraically solve for $y'(r)$ and $A_t'(r)$ in terms of the integration constants $c$ and $d$,

\beq
\label{solution}
y' = \frac{c}{\sqrt{\N^2 r^6 + \frac{d^2}{(2\p\a')^2}-c^2}}, \qquad
A_t' = \frac{d/(2\p\a')^2}{\sqrt{\N^2 r^6 +
\frac{d^2}{(2\p\a')^2}-c^2}}.
\eeq

\begin{figure}
\center
\includegraphics[width=0.8\textwidth]{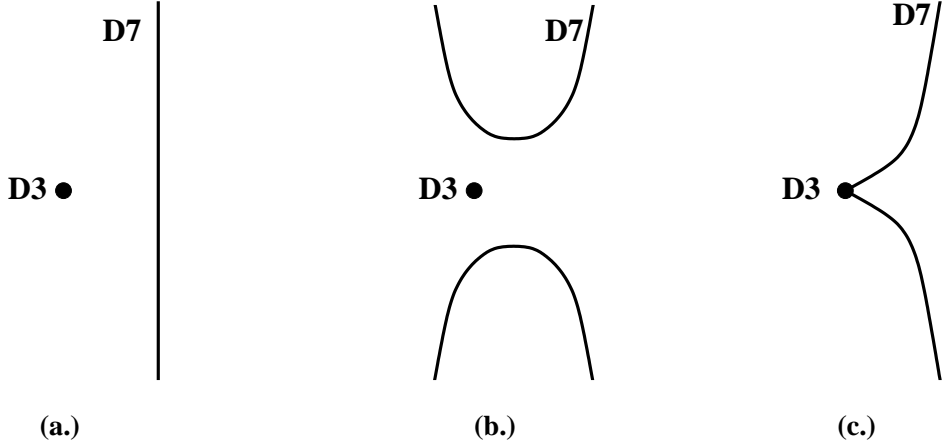}
\caption{\label{embeddings} Cartoons of the allowed D7-brane
embeddings at zero temperature. The vertical axis is one of the
directions transverse to the D3-branes but parallel to the
D7-branes. The horizontal axis is an overall transverse direction.
(a.) A Minkowski embedding corresponding to $y(r)$ being constant. (b.) A $\frac{d^2}{(2\p\a')^2} - c^2<0$
solution with turnaround describing a brane/anti-brane embedding as
in ref. \cite{Herzog:2007kh}. (c.) The zero-temperature
$\frac{d^2}{(2\p\a')^2} - c^2 >0$ horizon-crossing ``black hole''
embedding. }
\end{figure}

\noi These can be integrated using incomplete Beta functions (whose
properties we review in the Appendix). The result depends on the
sign of $\frac{d^2}{(2\p\a')^2}-c^2$. When $c=d=0$, we obtain the
solution with $y'(r)=0$ and $A_t'(r)=0$, so $y(r)$ and $A_t(r)$ are
constants. If $A_t(r)=0$, we recover the original embedding of ref.
\cite{KarchKatz} for a single straight D7-brane. This embedding is depicted in figure
(\ref{embeddings} a.). Solutions with $\frac{d^2}{(2\p\a')^2}-c^2$
positive and negative are also depicted in figure
(\ref{embeddings}). We address each of these in detail in the
following sections.

The action evaluated on these solutions is

\beq
\label{regaction}
S_{D7} = - \N \int^{\Lambda} dr r^3 \sqrt{\frac{\N^2 r^6}{\N^2 r^6 + \frac{d^2}{(2\p\a')^2} - c^2}}
\eeq

\noi The lower endpoint of integration will depend on the sign of
$\frac{d^2}{(2\p\a')^2} - c^2$. The integral diverges if we
integrate to $r = \infty$, so we regulate the integral with a cutoff
at $r = \Lambda$. The divergence is present for $d = c = 0$,
\textit{i.e.} for straight D7-branes with $A_t(r)=0$,

\beq
S_0 = - \N \int_0^{\Lambda} dr r^3 = - \frac{1}{4} \N \Lambda^4
\eeq

\noi In what follows we will obtain the renormalized on-shell
action, $S_{ren}$, by subtracting this divergence,

\beq
S_{ren} = \lim_{\Lambda \rarrow \infty} (S_{D7} - S_0)
\eeq

In the SYM theory, the thermodynamic potential $\Omega$ of the grand
canonical ensemble is given by $\Omega = - S_{ren}$. We Legendre transform to obtain the free energy density in the canonical ensemble, $F = \Omega + \mu d$. Recall that at zero temperature, free energy
and energy are the same, so at zero temperature $F$ is also the
energy density. The conserved charges $c$ and $d$ determine $\la \Om
\ra$ and $\la J^t \ra$ as follows,

\beq
\la \Om \ra = \frac{\delta \Omega}{\delta m} = - (2\p\a')
\frac{\delta S_{D7}}{\delta y(\infty)}, \qquad \la J^t \ra = -
\frac{\delta \Omega}{\delta \mu} = \frac{\delta S_{D7}}{\delta
A_t(\infty)}
\eeq

\noi where in each case when we vary one field we hold the other fixed. We then have

\beq
\delta S_{D7} = \int dr \left ( \frac{\delta L}{\delta A_t'(r)}
\partial_r \delta A_t(r) + \frac{\delta L}{\delta y'(r)} \partial_r
\delta y(r) \right) = d \delta A_t(\infty) - c \delta y (\infty)
\eeq

\noi where we demand that $\delta A_t(r)$ and $\delta y(r)$ are always zero at the lower endpoint of the $r$
integration. If we vary $A_t(r)$ while holding $y(r)$ fixed ($\delta y(r) = 0$), we find
$\la J^t \ra = d$, and similarly we find $\la \Om \ra = (2\p\a') c$.
Starting now, we will stick to the notation $c$ and $d$ and refer to these as the
condensate and density.

Using $\Omega = - S_{ren}$, we can determine whether solutions with
nonzero $d$ and $c$ are thermodynamically favored relative to the
single straight D7-brane with zero gauge field. $S_{ren} >0$ means
the configuration is \textit{favored} and $S_{ren} < 0$ means the
configuration is \textit{disfavored}.

\subsection{Brane/Anti-Brane Embeddings}  \label{ddbar}

We begin with $\frac{d^2}{(2\p\a')^2}- c^2 < 0$. In particular, we will begin with $d=0$ solutions, as analyzed recently in ref. \cite{Herzog:2007kh}. For such $d=0$ embeddings, $y'(r)$ diverges as $\frac{1}{\sqrt{r-r_0}}$ at
a critical radius $r_0 = \N^{-1/3} c^{1/3}$. This behavior indicates
that the D7-brane has a turn-around point where the D7-brane
smoothly matches onto a second branch. This solution thus describes
a D7/anti-D7-brane pair connected by a smooth neck, as depicted
schematically in figure (\ref{embeddings} b.).

In the asymptotic region, we may specify the distances of the
D7-brane and the anti-D7-brane to the D3-branes separately. In other
words, we may associate two masses to such a configuration. We
denote these as $m$ and $\bar{m}$. A D7-brane solution exists for
any $m$ and $\la \Om \ra$. The mass (or asymptotic separation) of
the anti-D7-brane is then fixed in terms of these two input
parameters. From the field theory point of view, we would like to
interpret the two masses $m$ and $\bar{m}$ as input parameters, so
that the above construction dynamically determines the value of the
condensate $\la \Om \ra$. From the supergravity picture, we can see that the condensate only depends on
the difference of the masses, $m - \bar{m}$. This is another
reflection of the fact that the embedding equation lost all
knowledge of the warp factors, and effectively we are solving for
flat-space embeddings. Only the relative position of the D7-branes
and anti-D7-branes matters, not their distance from the origin. As noted in ref. \cite{Herzog:2007kh}, however, all the connected D7/anti-D7-brane solutions have higher energy than the straight
D7/anti-D7-brane pair, and hence will be meta-stable, at best. The
true ground state of the system has vanishing $\la \Om \ra$.

Now we introduce nonzero $d$. To connect the two halves of the D7-brane smoothly we need the gauge field and its derivative to be continuous at the neck. In particular, the field strength on the two branches should be equal in magnitude but opposite in direction. The baryon number density is then equal and opposite at the two ends, so this configuration does not describe any net baryon number. The Lagrangian has a global $U(1) \times U(1)$ symmetry, and, at least in the grand canonical ensemble, we can imagine turning on different chemical potentials for the two $U(1)$'s at the level of the Lagrangian, even though the symmetry is spontaneously broken to the diagonal $U(1)_B$ in the state represented by the connected D7-brane configuration. If we wanted to have a brane/anti-brane system with finite baryon density we would have to include an explicit source at the neck. Such additional sources will back-react and alter the configuration, so we will not discuss this option further.

In any case, these brane/anti-brane systems correspond to a different Lagrangian than that for systems with a single D7-brane reaching the asymptotic region. The dual field theory for the brane/anti-brane is ${\cal N}=4$ SYM coupled to two hypermultiplets that preserve opposite ${\cal N}=2$ supersymmetries, so that the full theory is non-supersymmetric. When comparing all possible bulk configurations that correspond to hypermultiplets preserving $\N=2$ supersymmetry, the brane/anti-brane configurations never contribute. We will therefore only compare these solutions to the straight D7-brane and anti-D7-brane.

For nonzero $d$, keeping in mind that for the brane/anti-brane system this does not correspond to
a net baryon density, the D7-brane reaches the turn-around point

\beq
r_0 = \N^{-1/3} \left( c^2 - \frac{d^2}{(2\p\a')^2} \right)^{1/6}
\eeq

\noi The integrals for $y(r)$, $A_t(r)$ and the regulated on-shell action are performed in the Appendix. We find

\beq
y(r) = c \frac{1}{6} \N^{-1/3} \left( c^2 - \frac{d^2}{(2\p\a')^2} \right)^{-1/3} \left ( B\left( \frac{1}{3},\frac{1}{2} \right) - B\left( \frac{r_0^6}{r^6}; \frac{1}{3},\frac{1}{2} \right) \right )
\eeq

\noi and $A_t(r)$ is simply $\frac{d}{(2\p\a')^2} \frac{1}{c}$ times $y(r)$, as shown in eq. (\ref{ratio}). Notice $y(r_0)= A_t(r_0) = 0$. The mass and chemical potential are given by the asymptotic values at $r \rarrow \infty$, with $\lim_{r \rarrow \infty} B\left(\frac{r_0^6}{r^6}; \frac{1}{3},\frac{1}{2} \right) = 0$. The
renormalized on-shell action for the D7-brane is

\beq
S_{ren} =  - \frac{1}{6} \N^{-1/3} \left( c^2 -
\frac{d^2}{(2\p\a')^2} \right)^{2/3} B\left(-\frac{2}{3},\frac{1}{2}
\right)
\eeq

\noi The action for the anti-D7-brane is identical, so the total action is twice $S_{ren}$. Notice that $S_{ren} < 0$, so these connected brane/anti-brane configurations always have a higher free energy than the straight brane and anti-brane with constant $A_t(r)$, and hence are thermodynamically disfavored, just as in the $d=0$ case.

At this point we may also consider $c = \frac{d}{2\pi\alpha'} \neq
0$ so that $\frac{d^2}{(2\pi\alpha')^2} - c^2 = 0$. This
configuration shares a similar feature with the brane/anti-brane
configurations: it does not satisfy the required UV boundary
conditions for a theory with a single D7-brane added. When
$\frac{d^2}{(2\pi\alpha')^2} -c^2 = 0$, we have $y'(r)
\sim\frac{1}{r^3}$, so the D7-brane misses the D3-branes at the
origin and returns to the asymptotic region at $r \rightarrow
\infty$. We will therefore not study the $c = \frac{d}{2\pi\alpha'}$
case further.

\subsection{Finite-Density Black Hole Embeddings}  \label{bhemb}

We now consider $\frac{d^2}{(2\p\a')^2}- c^2 > 0$. In this section
we will show that these are the zero-temperature limit of black hole
embeddings. First, notice that these solutions have no turn-around
point. They describe D7-branes alone.

For a $\frac{d^2}{(2\p\a')^2}- c^2 >0$ solution, suppose we take
$c=0$. The solution with constant $y(r)$ is still allowed.
Geometrically, this is the zero-temperature Minkowski embedding, but
now with nonzero $A_t'(r)$. As mentioned before, the constant charge
density along the D7-brane requires a source at $r=0$, so this
embedding is not physical unless we add extra strings connecting the
D3-branes and D7-branes, or some other source. We will return to
these $c=0$ embeddings in the next subsection.

We can avoid this issue, however, by demanding that the D7-brane
touch the horizon at $r=0$. Since the horizon is located at $0=r_6^2
= r^2 +y^2$, this immediately implies a boundary condition on $y$

\beq
y(r=0) = 0.
\label{bc}
\eeq

\noi The embedding satisfying this boundary condition is the zero
temperature analogue of black hole embeddings. We will stick to this
name, even though at zero temperature ``horizon crossing'' would be
more appropriate. Using the change of variables $\tan \theta = y/r$,
we see that the slope of $y(r)$ at $r=0$ sets the value of $\theta$ at the horizon. This black hole embedding is
displayed schematically in figure (\ref{embeddings} c.).

The integral for $y'(r)$ in eq.(\ref{solution}) is done in the Appendix, with the result

\beq
y(r) = c \frac{1}{6} \N^{-1/3} \left( \frac{d^2}{(2\p\a')^2} - c^2 \right)^{-1/3} B\left( \frac{\N^2 r^6}{\N^2 r^6 + \frac{d^2}{(2\p\a')^2} - c^2}; \frac{1}{6},\frac{1}{3} \right)
\eeq

\noi and again, $A_t(r)$ is $\frac{d}{(2\p\a')^2} \frac{1}{c}$ times
$y(r)$. Notice that $y(0) = A_t(0)=0$. Identifying $m$ and $\mu$
from the asymptotic values $y(\infty)$ and $A_t(\infty)$, where as $r \rarrow \infty$ the
incomplete Beta function becomes $B\left(\frac{1}{6},\frac{1}{3}
\right)$, we find

\begin{subequations}
\label{relations}
\beq
c = \gamma \N (2\p\a')^3 \left( \m^2 - m^2 \right) m
\eeq
\beq
\frac{d}{2\p\a'} = \gamma \N (2\p\a')^3 \left( \m^2 -
m^2 \right) \mu
\eeq
\end{subequations}

\noi where we have defined the constant

\beq
\gamma \equiv  \left( \frac{1}{6} B\left(\frac{1}{6},\frac{1}{3}\right)
\right)^{-3} \approx 0.363 .
\eeq

Note that since these black hole embeddings were valid only when
$\frac{d^2}{(2\p\a')^2}- c^2 >0$, we can immediately see that they
only exist for $\mu > m$. In a theory with a mass gap, chemical
potentials of magnitude less than the mass gap are basically
trivial: they do not lead to any finite density. In the canonical
ensemble they never appear, since zero density corresponds to $\mu =
m$. In the grand canonical ensemble we should of course be able to
dial $\mu$ all the way down to zero, but the physics for $\mu < m$ is still expected to be trivial in
this range, with $d=0$ and the free energy being independent of
$\mu$. These expectations will be verified when we study
Minkowski embeddings in the next subsection. In this sense, the black hole embeddings cover
the whole range of physically interesting values of $\mu$.

The functional form of the condensate $c(\mu,m)$ is exactly that of
a charged boson in the presence of a chemical potential. The
chemical potential leads to a negative mass squared for the boson. Once this is larger than the positive mass squared in the Lagrangian, and if the scalar has a quartic coupling, the scalar will condense, its expectation value settling to the new minimum. Eqs. (\ref{relations}) can thus be viewed
as determining the effective quartic potential at strong coupling.
Notice that this also confirms the field theory expectation that at
finite density the ground state is described by scalar condensates,
not by a Fermi surface.

The renormalized on-shell action in this case is

\beq
S_{ren} = \frac{1}{4} \gamma^{-1/3} \N^{-1/3} \left(
\frac{d^2}{(2\p\a')^2} - c^2 \right)^{2/3} = \frac{1}{4} \gamma \N
(2\p\a')^4 \left(\mu^2-m^2 \right)^2 \label{bhrenaction}
\eeq

\noi Notice that $S_{ren} > 0$, so these embeddings are
thermodynamically favored relative to the straight D7-brane with no
gauge field. In the next subsection we will compare to the straight
D7-brane with a nonzero gauge field. Notice that we observe no
instability in $\Omega$. Stability requires $\frac{\partial d}{\partial\mu} \geq
0$, which is clearly satisfied in eq. (\ref{relations}). This is
consistent with the numerical results of ref.
\cite{Kobayashi:2006sb} in the zero-temperature limit.

The thermodynamic potential in the canonical ensemble is

\beq
F = \Omega + \mu d = \frac{1}{4} \N^{-1/3} \gamma^{-1/3} \left( \frac{d^2}{(2\p\a')^2} - c^2 \right)^{-1/3} \left(3\frac{d^2}{(2\p\a')^2} + c^2 \right)
\eeq

\noi so that $F > 0$. We will see in the next subsection that the black hole embedding still turns out to be the configuration with lowest $F$, however.

\subsection{Finite-Density Minkowski Embeddings}  \label{minkemb}

Let us return to $c=0$ and constant $y(r)$. These are solutions
everywhere away from $r=0$. These are straight D7-branes, which do not obey the
boundary condition $y(0)=0$. In other words, when $r=0$ the
D7-branes are still a distance $y$ away from the D3-branes. The mass
$m$ is given by $m = \frac{y}{2\p\a'}$.

The integral for $A_t(r)$ will be unchanged, however. In particular,
we still integrate from $r=0$ to $r=\infty$, and $A_t(0)=0$. From
eq. (\ref{relations}) we read off the relation between $\mu$ and
$d$,

\beq
d = \gamma \N (2\p\a')^4 \mu^3,
\eeq

\noi The integral for the on-shell action is also unchanged, so we
simply set $c=0$ in the first equality of eq. (\ref{bhrenaction}),

\beq
S_{ren}  = \frac{1}{4} \gamma^{-1/3} \N^{-1/3} (2\p\a')^{-4/3} d^{4/3} = \frac{1}{4} \gamma \N (2\p\a')^4 \mu^4
\eeq

\noi and hence $\Omega = -S_{ren} \sim - \mu^4$. Naively this
$\Omega$, for any nonzero $d$ (or $\mu > m$), is smaller (more
negative) than the one for the black hole embedding, eq.
(\ref{bhrenaction}). Notice, however, that unlike the nice
interpretation we found for the relations among $c$, $d$, $\mu$ and $m$
for the black hole embeddings, this result for $\Omega$, a pure
quartic in $\mu$, is completely independent of the mass and
therefore appears to be unphysical. For example, in the decoupling
limit $m \rarrow \infty$, $\Omega$ should go to zero.

The problem of course is that the Minkowski embedding, as it stands,
is not consistent: we completely ignored the contribution to
$\Omega$ due to whatever object sources the D7-brane gauge field.
The simplest source is a finite density $d$ of fundamental strings
stretching from the D3-branes to the D7-brane. These D7-branes with strings attached are not solutions
to the full equations of motion following from the combined DBI and Nambu-Goto actions, however.
Naively, we might think that the string back-reaction on the D7-brane can be neglected in the large-$N_c$ and large-$\lambda$ limit. For a \textit{single} string this is certainly true. The prefactor of the Nambu-Goto action scales as $\sqrt{\lambda}$, while from eq. (\ref{dbiprefactor}) we see that the prefactor $\N$ of the DBI action scales as $N_f N_c \lambda$. We want a finite density $d$ of strings, however. If we keep all geometric distances of order one in units of the AdS curvature radius, the mass will scale as $\sqrt{\lambda}$, so we want the chemical
potential also to scale as $\sqrt{\lambda}$. From eq. (\ref{relations}), we can
see that the density $d$ will then scale as $\Nf \Nc \sqrt{\lambda}$. With this, $d$ times the Nambu-Goto action has the same $N_f N_c \lambda$ scaling as the DBI action, and so the back-reaction is order one.

To include the back-reaction one has to re-solve the equations of motion including the extra source term. We already found
the most general solution to the equations of motion with sources localized at $r=0$, so the back-reacted solution including any such sources must be within this class. The only well-behaved solution in this class which asymptotically becomes a single brane is the black hole embedding, so it must be the back-reacted solution.

To show that this is physically reasonable we want to show that the brane-plus-strings configuration, which
could serve as a consistent initial data for a full time-dependent physical solution, has higher energy than the black hole embedding. We add a term $d m$ to the action representing the finite density $d$ of strings with length $ y = (2 \pi \alpha') m$. The full free energy including the D7-branes and strings in this case is

\beq \frac{\Omega}{ \frac{1}{4} \gamma \N (2 \pi \alpha')^4} = - \mu^4 + 4 \mu^3 m = \mu^3 (4
m - \mu)
\eeq

\noi which is positive at $\mu=m$ and is disfavored relative to the black hole embedding at all values of $\mu>m$. We can also allow some mixture, where the D7-brane embedding satisfies the boundary condition
$y(r=0)=y_0$ for some $m > \frac{y_0}{2 \pi \alpha'} > 0$, with a density $d$ of strings extending from $y=0$ to $y=y_0$. In figure (\ref{embeddings}), we imagine sliding the black hole embedding to the right a distance $y_0$, such that the total asymptotic separation remains $y$, and then introducing the strings. The free energy is then

\beq \frac{\Omega}{\frac{1}{4} \gamma \N (2 \pi \alpha')^4} =
 - \left( \mu^2 - \left( m - \frac{y_0}{2 \pi \alpha'} \right)^2 \right)^2 + 4 \mu \left( \mu^2 - \left( m - \frac{y_0}{2 \pi \alpha'} \right)^2 \right) \frac{y_0}{2 \pi \alpha'}
\eeq

\noi and we can easily check that in the relevant regime $\mu>m >
\frac{y_0}{2 \pi \alpha'} > 0$ the derivative of this expression
with respect to $y_0$ is strictly positive. The D7-brane with
strings attached can continuously lower its energy until it turns
into the black hole embedding with all strings dissolved in the D7-brane.

While none of the finite-density Minkowski embeddings solve the equations of motion, the trivial Minkwoski embeddings with constant $A_t(r)$ do. As only $F_{rt}$ enters the action and not $A_t(r)$ itself, these embeddings have $F=\Omega=d=0$, and $\mu$ is a free parameter. For $\mu<m$ these are the only allowed configurations, so they dominate the ensemble. The fact that they are indistinguishable from the vacuum is completely natural from the field theory point of view: the lightest charge carriers have mass $m$, and for $\mu <m$ no non-zero density can be produced. When $\mu >m$ these solutions still exist, but in that regime the black hole embedding, with negative $\Omega$, dominates the ensemble. At $\mu=m$ we hence have a second order phase transition between the trivial Minkowski embedding and the
 black hole embeddings. Both $\Omega$ and its first derivative with respect to $\mu$ are zero at that point, but the second derivative jumps from zero for the trivial Minkowski embedding to $-2 \gamma \N (2\p\a')^4 m^2$ for the black hole embedding. Again, this is exactly what we expect for a bosonic condensate with a non-trivial effective quartic interaction. The critical exponents are given by their mean-field values, that is, for $\mu = m + \epsilon$ we have $\Omega \sim \epsilon^2$, $c \sim \epsilon$, $d \sim \epsilon$.

In the canonical ensemble, the trivial Minkowski embeddings do not even contribute because they have $d=0$. The finite-density Minkowski embeddings with strings attached that we considered above are not solutions of the equations of motion, so they cannot contribute. We thus conclude that, in the canonical ensemble, only black hole embeddings contribute. This analytically confirms the results of ref. \cite{Kobayashi:2006sb}, at least at zero temperature.

Lastly, notice that the configuration we mentioned above of a single
straight string stretching between a D3-brane and a D7-brane in flat
space is BPS, preserving half of the supersymmetry. We should ask
why a density of such strings is not also BPS, and hence the
dominant configuration? From a field theory point of view, the
single string represents a single fundamental-representation
``quark.'' As the force between colored charges is Coulombic, we may
introduce a single quark into the system and still make the total
state of the system gauge invariant by placing an anti-fundamental
color charge at infinity. A finite density of such states, as would
be represented by the configuration we studied above, is not a
physical configuration, however, because it has a finite color
density. It does not represent a gauge invariant state that
contributes to the partition function. Both at strong and weak
coupling a scalar condensate forms at finite density, breaking the
$SU(N_c)$ gauge group to $SU(N_c-1) \times U(1)$. In other words,
the system moves out onto the Higgs branch. No BPS states exist on
the Higgs branch. For example, a finite density of baryons is gauge
invariant but is not BPS. Our analysis is consistent with this field
theory picture: the D7-brane-with-strings configuration is not a
supergravity solution, and in the SYM theory we indeed find that a
scalar condensate forms.

\subsection{General Zero-temperature Probe D-brane Embeddings}  \label{general}

While the discussion so far was only for the D3/D7 system and the
only field strength we turned on was $F_{rt}$, corresponding to
finite density in the field theory, the same methods can be used to
generate analytic embeddings for a variety of probe D-brane systems
and worldvolume fields. In this subsection we will consider
D$q$-brane probes in the D3-brane background, with $q = 5,7$. The
probe D5-brane wraps $AdS_4 \times S^2$ inside $AdS_5 \times S^5$,
and corresponds to adding flavor fields confined to a
(2+1)-dimensional defect in the SYM theory
\cite{Karch:2000gx,DeWolfe:2001pq}.

We include $F_{rt}$ and all possible gauge field components $F_{ri}$ with spatial indices $i$ parallel to the D3-branes, corresponding to non-vanishing $U(1)_B$ current components $\la J_i\ra$ in the field theory. We are working at zero temperature, so these currents will not dissipate and can be turned on as constant background fields. They do not require an electric field to be turned on. For all these cases, the warp factors continue to cancel in the DBI action, and we can obtain analytic solutions for the embeddings that basically correspond to D$q$-brane embeddings in flat space. For the interesting case of non-vanishing electric or magnetic fields, the warp factors do not cancel and one has to solve the full-fledged non-linear problem in curved space.

Our general D$q$-brane will be extended along $AdS_{D+2} \times S^n$, so that $i$ runs from one to $D$. Generalizing eq. (\ref{dbi}), the action for the D$q$-brane is

\beq
\label{generaldbi}
S_{Dq} = - N_f T_{Dq} \int d^{q+1} \xi \sqrt{-det (g_{ab} + (2\p\a') F_{ab})}
\eeq

\noi where $T_{Dq}$ is the tension of the D$q$-brane. Introducing $y(r)$, $A_t(r)$ and $F_{ri} = A_i'(r)$, the Lagrangian for the D$q$-brane is

\beq
L = -\N_q r^n \sqrt{1+ y'^2 - (2\p\a')^2 A_t'^2 + (2\p\a')^2 \sum_{i} A_i'^2}
\eeq

\noi where $\N_q = \Nf T_{Dq} V_n$ with $V_n$ the volume of a unit $n$-sphere. We will divide both sides of eq. (\ref{generaldbi}) by the volume of ${\mathbb{R}}^{D,1}$ so that $S_{Dq} = \int dr L$. To simplify the formulas let us introduce the following notation:

\beq
\vec{Y} = \left( y,(2\p\a')A_t,(2\p\a')A_i \right), \qquad \vec{C} = \left( c, \frac{d}{(2\p\a')},\frac{J_i}{(2\p\a')} \right)
\eeq

\noi that is, we construct a vector of fields and the associated conserved charges, with components $Y_{a}$ and $C_{a}$. The inner product of these vectors has a Lorentzian signature, \textit{i.e.} $diag(+1,-1,+1,+1,+1)$ for the D7-brane. With this, we have

\beq
L = -\N_q r^n \sqrt{1+ Y^{'2}}
\eeq

\noi and from this we get

\beq
C_a = \frac{\N_q r^n Y^{'}_a}{\sqrt{1+Y^{'2}}}
\eeq

\noi We algebraically solve for the fields, and find the regulated on-shell action,

\beq
Y_a^{'2} = \frac{C_a^2}{\N_q^2 r^{2n} - C^2}, \qquad S_{Dq} = - \N_q \int^{\Lambda} dr r^n \sqrt{\frac{\N_q^2 r^{2n}}{\N_q^2 r^{2n} - C^2}}.
\eeq

\noi Once again, the lower endpoint of integration depends on the sign of $C^2$. The divergence in $S_{Dq}$ is cancelled by the action for a single straight D$q$-brane with zero field strength,

\beq
S_0 = -\N_q \int_0^{\Lambda} dr r^n = - \frac{\N_q}{n+1} \Lambda^{n+1}, \qquad S_{ren} = \lim_{\Lambda \rarrow \infty} \left( S_{Dq} - S_0 \right).
\eeq

$\frac{d^2}{(2\p\a')^2} - c^2 < 0$ for the D7-brane becomes $C^2 > 0$ for the D$q$-brane. The D$q$-brane has a turn-around point at $r_0 = \left(\frac{C^2}{\N_q^2}\right)^{\frac{1}{2n}}$. The integrals can be done using the machinery in the Appendix by changing coordinates to $t = r_0^{2n}/r^{2n}$,

\beq
Y_a = C_a \frac{1}{2n} \N_q^{\frac{-1}{n}} (C^2)^{\frac{1-n}{2n}} \left ( B\left( \frac{n-1}{2n},\frac{1}{2}
\right) -B\left( \frac{r_0^{2n}}{r^{2n}}; \frac{n-1}{2n},\frac{1}{2}
\right) \right ),
\eeq

\beq
S_{ren} = - \frac{1}{2n} \N_q^{\frac{-1}{n}} \left(C^2\right)^{\frac{1+n}{2n}} B\left( -\frac{n+1}{2n}, \frac{1}{2} \right)
\eeq

\noi Notice $S_{ren} < 0$, so the free energy is positive and hence these embeddings are thermodynamically disfavored
relative to the embedding with a straight brane/anti-brane pair (which still has $\Omega=0$), just as in section \ref{ddbar}.

$\frac{d^2}{(2\p\a')^2} - c^2 > 0$ for the D7-brane becomes $C^2 < 0$ for the D$q$-brane. The D$q$-brane does not have a turn-around point and the integrations extend to $r=0$. Here the change of variables is to $u = \frac{\N_q^2}{|C^2|} r^{2n}$, with the result

\beq
Y_a = C_a \frac{1}{2n} \N_q^{-\frac{1}{n}} |C^2|^{\frac{1-n}{2n}} B \left ( \frac{\N_q^2 r^{2n}}{\N_q^2 r^{2n} + |C^2|};\frac{1}{2n},\frac{n-1}{2n} \right )
\eeq

\beq
S_{ren} = - \frac{1}{2n} \N_q^{-\frac{1}{n}} |C^2|^{\frac{1+n}{2n}} B \left ( \frac{2n+1}{2n}, - \frac{n+1}{2n} \right)
\eeq

\noi As in the $\frac{d^2}{(2\p\a')^2} - c^2 > 0$ case for the D7-brane, $S_{ren} > 0$, so the free energy is negative and these embeddings are thermodynamically favored.

\section{The Limit of Zero Mass}  \label{zeromass}

In this section we study the SYM theory at finite temperature and density, and take $m=0$. At finite temperature, the $AdS_5$ metric in eq. (\ref{kkmetric}) becomes the AdS-Schwarzschild metric,

\beq
ds^2 = \frac{dr_6^2}{f(r_6)} - f(r_6)dt^2 + r_6^2 d\vec{x}^2 + d\theta^2 + \sin^2\theta ds_{S^1}^2 + \cos^2 \theta ds_{S^3}^2
\eeq

\noi where $f(r_6) = r_6^2 - \frac{r_H^4}{r_6^2}$. The black hole horizon is related to the temperature by $r_H = \pi T$. The D7-brane embedding is described by $\theta(r_6)$. Introducing the worldvolume gauge field $F_{r_6 t} = \partial_{r_6} A_t(r_6)$, the D7-brane action eq. (\ref{dbi}) becomes

\beq
S_{D7} = -\N \int^{\Lambda} dr_6 r_6^3 \cos^3 \theta \sqrt{1 + f(r_6) \theta'^2 - (2\p\a')^2 A_t'^2}
\label{finiteTdbi}
\eeq

\noi where we have divided by the volume of ${\mathbb{R}}^{3,1}$, $\N$ is the same as defined above, and primes denote $\frac{\partial}{\partial r_6}$. At this point we can plainly see that the system has only one conserved charge,

\beq
d \equiv \N r_6^3 \cos^3 \theta \frac{(2\p\a')^2A_t'}{\sqrt{1 + f(r_6) \theta'^2 - (2\p\a')^2 A_t'^2}}
\eeq

\noi The algebraic solution for $A_t'(r_6)$ is

\beq
A_t'(r_6) = \frac{d}{(2\p\a')^2} \sqrt{\frac{1 + f(r_6) \theta'^2}{\N^2 r_6^6 \cos^6 \theta + \frac{d^2}{(2\p\a')^2}}}.
\label{finiteTgaugesol}
\eeq

\noi Inserting this into $S_{D7}$, we find the on-shell action,

\beq
S_{D7} = - \N^2 \int^{\Lambda} dr_6 r_6^6 \cos^6 \theta \sqrt{\frac{1 + f(r_6) \theta'^2}{\N^2 r_6^6 \cos^6 \theta + \frac{d^2}{(2\p\a')^2}}}
\label{finiteTonshelldbi}
\eeq

\noi We may obtain the equation of motion for $\theta(r_6)$ in two ways. We may either derive it from eq. (\ref{finiteTdbi}) and then plug in the gauge field eq. (\ref{finiteTgaugesol}), or we may eliminate the gauge field at the level of the action via a Legendre transform and then derive $\theta(r_6)$'s equation of motion. The Legendre-transformed action, $\bar{S}_{D7}$, is

\beq
\label{lteffaction}
\bar{S}_{D7} = S_{D7} - \int dr_6 F_{r_6 t} \frac{\delta S_{D7}}{\delta F_{r_6 t}} = - \int dr_6 \sqrt{1 + f(r_6) \theta'^2} \sqrt{\N^2 r_6^6 \cos^6 \theta + \frac{d^2}{(2\p\a')^2}}.
\eeq

\noi Once again, to cancel the $\Lambda \rarrow \infty$ divergence in $S_{D7}$ we subtract the action of a single straight D7-brane with no gauge field. At finite temperature, the lower endpoint of integration is now $r_H$, hence

\beq
S_0 = - \N \int^{\Lambda}_{r_H} dr_6 r_6^3 = - \frac{1}{4} \N \Lambda^4 + \frac{1}{4} \N r_H^4,
\eeq

\noi and $S_{ren}$ is defined as above.

A valid solution of $\theta(r_6)$'s equation of motion is simply $\theta(r_6) = 0$, corresponding to $m=0$ and $\la \Om \ra = 0$ in the SYM theory. For this solution, the integrals for $A_t(r)$ and $S_{D7}$ are identical to those for a black hole embedding with $c=0$, eqs. (\ref{solution}) and (\ref{regaction}), but with $r \rarrow r_6$ and with the lower endpoint of integration $r_H$. We thus find

\beq
A_t(r_6) = \frac{1}{6} \N^{-1/3} (2\p\alpha')^{-4/3} d^{1/3} \left( B\left( \frac{\N^2 r_6^6}{\N^2 r_6^6 + \frac{d^2}{(2\p\a')^2}}; \frac{1}{6}, \frac{1}{3} \right) - B\left( \frac{\N^2 r_H^6}{\N^2 r_H^6 + \frac{d^2}{(2\p\a')^2}}; \frac{1}{6}, \frac{1}{3} \right)  \right)
\eeq

\noi Notice that $A_t(r_H) = 0$. The value of $\mu$ is given by taking $r_6 \rarrow \infty$, for which the first term in parentheses goes to $B\left( \frac{1}{6}, \frac{1}{3} \right)$. For the on-shell action we find

\beq
S_{ren} = - \frac{1}{4} \N r_H^4 - \frac{1}{6} \N^{-1/3} \left(\frac{d}{2\p\a'}\right)^{4/3} B \left ( \frac{\frac{d^2}{(2\p\a')^2}}{\N^2 r_H^6 + \frac{d^2}{(2\p\a')^2}}; - \frac{2}{3}, \frac{7}{6} \right ).
\eeq

\noi The $T = 0$ limit of course reproduces the results of section \ref{bhemb} for black hole embeddings with $c=0$. A straightforward analysis reveals that $\Omega = -S_{ren}$ has no meaningful features: in the $m=0$ limit, the system is stable and exhibits no phase transitions.

\section{Conclusion}  \label{conclusion}

For $\Nf$ massive $\Ntwo$ supersymmetric hypermultiplet fields at
finite $U(1)_B$ density coupled to the $\N=4$ SYM theory in the
limits of large-$\Nc$ and large $\lambda$, we have provided analytic
expressions for the hypermultiplets' leading-order (in $\Nf/\Nc$)
contribution to the free energy in two limits: zero temperature and
zero mass. We have found a second order transition at zero
temperature when $\mu =m$. In the supergravity description, we have
confirmed the proposal of refs.
\cite{Kobayashi:2006sb,Ghoroku:2007re} that, in the canonical
ensemble, only black hole D7-brane embeddings are physical, while in
the grand canonical ensemble black hole embeddings are appropriate
for $\mu > m$ while Minkowski embeddings with constant gauge field
(and zero field strength) are appropriate when $\mu < m$. Perhaps
the most obvious task for the future is a thorough analysis of the
phase diagram of this theory in the grand canonical ensemble.

\section*{Acknowledgments}

We would like to thank C.~Herzog for useful discussions. The work of A.K. was supported in part by the U.S. Department of Energy under Grant No.~DE-FG02-96ER40956. The work of A.O'B. was supported in part by the Jack Kent Cooke Foundation.

\section*{Appendix: Beta functions and Incomplete Beta functions}
\setcounter{equation}{0}
\renewcommand{\theequation}{\arabic{equation}}

The integrals needed in the bulk of the text all take the form of Beta functions or incomplete Beta functions. For the reader's convinience we collect some of the important properties of these functions in this appendix. The Beta function is defined as

\beq
\label{bf}
B(a,b) = \frac{\G(a)\G(b)}{\G(a+b)} = \int_0^1 dt (1-t)^{b-1} t^{a-1} = \int_{0}^{\infty} du (1+u)^{-(a+b)} u^{a-1}
\eeq

\noi and the incomplete Beta function as

\beq
B\left( x;a,b \right) = \int_0^{x} dt (1-t)^{b-1} t^{a-1} = \int_{0}^{x/(1-x)} du (1+u)^{-(a+b)} u^{a-1}.
\label{ibf}
\eeq

\noi These satisfy the recursion relation

\beq
B(x;a,b) = B(a,b) - B(1-x;b,a)
\eeq

\noi and are related to the hypergeometric function as

\beq
B(x;a,b) = a^{-1} x^a F(a,1-b;a+1;x).
\eeq

\noi The $\Lambda^4$ divergence in $S_{D7}$ appears in the expansion, for $a = -2/3$ about $x=0$,

\beq
B\left( x; - \frac{2}{3}, b \right) = - \frac{3}{2} x^{-2/3} + O(x^{1/3})
\eeq

\noi where for $S_{D7}$ we will have $x \sim \Lambda^{-6}$ (see for example eq. (\ref{sless}) below).

The integrals we need can easily be brought into the form of eq. (\ref{ibf}). For notational simplicity, we will replace $\frac{d^2}{(2\p\a')^2}$ with just $d^2$. We begin with the $d^2 - c^2 < 0$ case of section \ref{ddbar}. The integral for $y(r)$ and $A_t(r)$, eq. (\ref{solution}), in this case is

\beq
I_<(r) = \int_{r_0}^{r} dr \left ( \N^2 r^6 - (c^2 - d^2) \right)^{-1/2},
\eeq

\noi with $r_0 = \N^{-1/3} \left( c^2 - d^2 \right)^{1/6}$. We change variables to $t = r_0^6 / r^6$,

\bea
I_<(r) & = & \frac{1}{6} \N^{-1/3} (c^2 - d^2)^{-1/3} \int_{t}^{1} dt (1-t)^{-1/2} t^{-2/3} \\ & = & \frac{1}{6} \N^{-1/3} (c^2 - d^2)^{-1/3} \left ( B\left( \frac{1}{3},\frac{1}{2} \right) -B\left( \frac{r_0^6}{r^6}; \frac{1}{3},\frac{1}{2} \right) \right ).\nonumber
\eea

\noi The same change of variables works for the regulated on-shell action eq. (\ref{regaction}),

\bea
\label{sless}
S_{D7}^< & = & - \frac{1}{6} \N r_0^4 \int_{r_0^6/\Lambda^6}^1 dt (1-t)^{-1/2} t^{-5/3} = - \frac{1}{6} \N r_0^4 \left ( B\left(-\frac{2}{3},\frac{1}{2} \right) - B\left(\frac{r_0^6}{\Lambda^6};-\frac{2}{3},\frac{1}{2}\right) \right ) \\ & = & - \frac{1}{6} \N r_0^4 B\left(-\frac{2}{3},\frac{1}{2} \right) - \frac{1}{4} \N \Lambda^4 + O(\Lambda^{-2}) \nonumber
\eea

\noi Now we turn to the $d^2 - c^2 > 0$ case of section \ref{bhemb}. The integral for $y(r)$ and $A_t(r)$ is

\beq
I_>(r) =\int_0^{r} dr \left ( \N^2 r^6 + d^2 - c^2 \right)^{-1/2}.
\eeq

\noi We change variables to $u = \frac{\N^2 r^6}{d^2 - c^2}$,

\bea
I_>(r) & = & \frac{1}{6}\N^{-1/3} \left( d^2 - c^2 \right)^{-1/3} \int_{0}^{u} du (1+u)^{-1/2} u^{-5/6} \\ & = & \frac{1}{6} \N^{-1/3} \left( d^2 - c^2 \right)^{-1/3} B\left( \frac{\N^2 r^6}{\N^2 r^6 + d^2 - c^2}; \frac{1}{6},\frac{1}{3} \right). \nonumber
\eea

\noi The regulated on-shell action is, with $\Lambda ' = \frac{\N^2 \Lambda^6}{d^2 - c^2}$

\bea
S_{D7}^> & = &- \frac{1}{6} \N^{-1/3} (d^2 - c^2)^{2/3} \int_0^{\Lambda'} du (1+u)^{-1/2} u^{1/6} \\ & = & - \frac{1}{6} \N^{-1/3} (d^2 - c^2)^{2/3} \left( B \left ( \frac{7}{6}, - \frac{2}{3} \right ) - B \left( \frac{1}{\Lambda' + 1}; - \frac{2}{3}, \frac{7}{6} \right ) \right ) \nonumber \\ & = & \nonumber - \frac{1}{6} \N^{-1/3} (d^2 - c^2)^{2/3} B \left ( \frac{7}{6}, - \frac{2}{3} \right ) - \frac{1}{4} \N \Lambda^4 +O(\Lambda^{-2}) \nonumber
\eea

\bibliography{gkp}
\bibliographystyle{JHEP}

\end{document}